# Infrared frequency combs and supercontinua for multiplex high sensitivity spectroscopy

J. Mandon[1], E. Sorokin[2], I.T Sorokina[2], G. Guelachvili[1], N. Picqué[1]

1. Laboratoire de Photophysique Moléculaire, CNRS, Université Paris-Sud, Bâtiment 350, 91405 Orsay Cedex, France
2. Institut für Photonik, TU Wien, Gusshausstrasse 27/387, A-1040 Vienna, Austria

**Abstract.** An infrared high-brightness light source based on supercontinuum generation through a SF6 photonic crystal fiber seeded by a $Cr^{4+}$:YAG femtosecond oscillator is developed for high resolution multiplex spectroscopy in the 1.5 µm region. Moreover, a multiplex high resolution approach based on a $Cr^{4+}$:YAG frequency comb enables to probe large spectral domains, with simultaneous sensitive measurement of the absorption and the dispersion associated with all individual spectral features.

## 1. INTRODUCTION

Femtosecond (fs) mode-locked lasers and supercontinua represent the first versatile table-top coherent sources linking broad bandwidth and high brightness. Femtosecond frequency combs offer additional new features for high resolution spectroscopy as they exhibit a broad spectrum, made of equidistant narrow lines of stable and accurate individual positions. Their comb structure may be taken advantage of for improved spectroscopy. This new generation of light sources requires new spectrometric methodologies to be developed.
In this article, our progress in the development of infrared supercontinua and frequency combs based methods is presented. Supercontinua are first being implemented for infrared absorption spectroscopy. A powerful method, femtosecond frequency comb spectrometry, is also being experimentally demonstrated. It is based on the exploitation of a frequency comb on its whole spectral domain is aimed at providing sensitivity, accuracy, resolution, broad spectral extension and rapid acquisition times. This multiplex approach enables to probe large spectral domains, with simultaneous measurement of the absorption and the dispersion associated with all individual spectral features.
Purposely a femtosecond $Cr^{4+}$:YAG frequency comb has been developed in the 1.5 µm range. In a previous paper [1], we used it as a bright source for high resolution absorption spectroscopy. Here, spectral broadening using highly non linear Photonic Crystal Fibers (PCF) is investigated. The $Cr^{4+}$:YAG oscillator is also applied to frequency comb spectrometry. As an illustration, the entire $\nu_1+\nu_3$ vibration-rotation band region of acetylene is recorded around 6500 cm$^{-1}$.

## 2. FEMTOSECOND $Cr^{4+}$:YAG OSCILLATOR BROADENED BY A PCF FIBER FOR BROADBAND SPECTROSCOPY

The femtosecond infrared oscillator, emitting in the 1.5 µm region, makes use of a 2 cm-long Brewster-cut $Cr^{4+}$:YAG crystal, pumped by a 1.06 µm commercial Nd:YVO4 laser. Stable mode-locking is obtained by a SEmiconductor Saturable Absorber Mirror and the dispersion is compensated by chirped mirrors. The output power is about 100 mW and the repetition rate of the pulses is 150 MHz. Stable operation is obtained in the positive dispersion regime, where 1.5 ps chirped pulses are obtained. They may be compressed down to 100 fs. As shown on Fig. 1, the laser output spectrum extends from 6340 to 6790 cm$^{-1}$ at 25 dB below the maximum. To broaden the available spectral domain, a highly non linear multimode PCF in SF6 glass, similar to the one described in [2,3], has been launched. Its core diameter is 4.5 µm, its length 40 cm and its non linear index is 2.2 x 10$^{-15}$ cm$^2$.W$^{-1}$. At the PCF output, the accessible spectrum extends over 2000 cm$^{-1}$ from 6010 to 7960 cm$^{-1}$ at 25 dB below the maximum, as shown on Fig. 1. This experimental set-up has been successfully used for the realization of high resolution spectra, as illustrated on the upper right part of Fig.1 which displays the region of the $\nu_2+\nu_3+\nu_4$ band of $NH_3$ at 0.12 cm$^{-1}$ resolution. Supercontinua prove to be a convenient and inexpensive way to perform broadband multiplex spectroscopy with table-top high brightness sources, as their spectral radiance is typically 10$^5$ higher than incoherent sources.

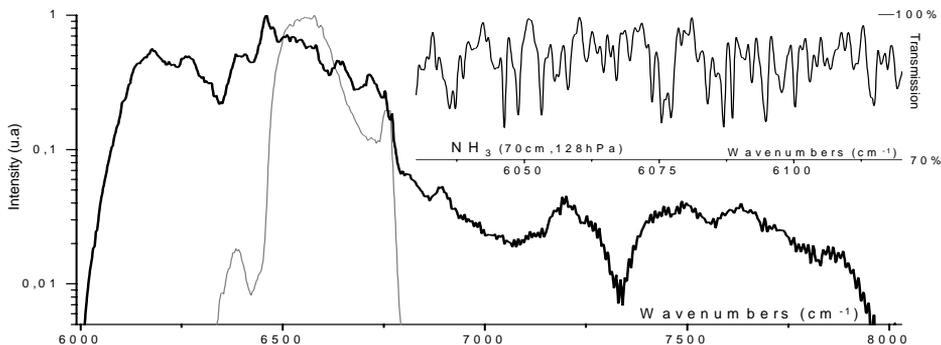

**Figure 1**. Low resolution spectra on a logarithmic vertical scale of the laser output (thin gray line) and the PCF output (bold black). Upper right part: portion of the ammonia spectrum recorded using the supercontinuum as an absorption source.

## 3. FEMTOSECOND FREQUENCY COMB SPECTROSCOPY

In order to enhance the sensitivity provided by broadband femtosecond light sources, a new method, femtosecond frequency comb spectroscopy, has been conceived [4] and its first demonstration is given here. The frequency comb probes a cell containing the molecular sample and is analyzed by a new kind of Fourier spectrometer characterized by the radio-frequency detection, at the comb repetition rate, of the interferogram. By performing in-phase and in-quadrature detection of the interferogram, the absorption and dispersion of the spectral features are measured simultaneously in the same experiment. Thanks to the synchronous detection performed in the radio-frequency domain, extremely high sensitivity is expected together with short acquisition time. Additionally, femtosecond frequency comb spectrometry provides broad spectral extension, high spectral resolution, and spatial resolution. The wavenumber scale may be self-calibrated by the frequency comb, with no need of molecular frequency standards. The method may be developed in any spectral domain.

As a first demonstration of our femtosecond frequency comb spectroscopy method, our homemade $Cr^{4+}$:YAG mode-locked femtosecond laser has been used as a broadband comb source. The laser beam passes through a 80-cm cell filled with acetylene. It is analyzed by the high resolution Connes-type interferometer of Laboratoire de Photophysique Moléculaire and detected by a fast InGaAs detector. Synchronous in-phase detection at 150 MHz frequency is performed by a lock-in amplifier. Figure 2 provides a restricted portion of the resulting absorption spectrum, exhibiting lines of the R-branch of the $\nu_1+\nu_3$ band of $C_2H_2$. Signal to noise ratio is still limited because of the absence of appropriate dynamics for the interferogram's measurement. This experiment however clearly validates the capabilities of our method and especially the good quality of the retrieved lineshapes. Improvement of the electronics is presently underway.

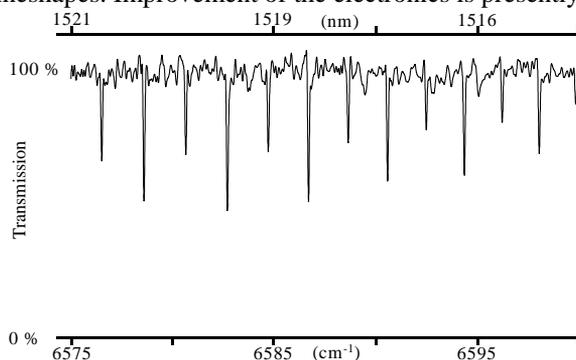

**Figure 2**. Portion of the overtone spectrum of the acetylene molecule (pressure: 59 hPa) in the 1.5 µm region obtained from femtosecond frequency comb spectroscopy.

**References**
[1] J. Mandon, G. Guelachvili, N. Picqué, F. Druon, P. Georges, Optics Letters **32**, 1677 (2007).
[2] V.L. Kalashnikov, E. Sorokin, S. Naumov, I.T. Sorokina, V.V. Ravi Kanth Kumar, A.K. George, Applied Physics B **79**, 591 (2004).
[3] V.V. Ravi Kumar, A. George, W. Reeves, J. Knight, P. Russell, F. Omenetto, A. Taylor, Optics Express **10**, 1520 (2002).
[4] N. Picqué, G. Guelachvili, Femtosecond frequency combs : new trends for Fourier transform spectroscopy, Fourier Transform Spectroscopy/Hyperspectral Imaging and Sounding of the Environment Topical Meetings on CD-ROM (The Optical Society of America, Washington, DC, 2005), paper FTuA2, 3 pages (2005).